\documentclass[showpacs,aps,prb,twocolumn,superscriptaddress]{revtex4-1}
\usepackage{graphicx}
\usepackage{subfigure}
\usepackage{bm}
\usepackage{color}
\usepackage{mathptmx}
\usepackage{amsmath}
\usepackage{amssymb}
\usepackage{enumerate}

\begin{document}

\title{Atomistic simulation of sub-nanosecond non-equilibrium field cooling processes  for magnetic data storage applications}
\author{R. F. L. Evans}
\email{richard.evans@york.ac.uk}
\affiliation{Department of Physics, University of York, York, YO10 5DD, UK}

\author{W. J. Fan}
\email{eleanorfan@163.com}
\affiliation{Department of Physics, University of York, York, YO10 5DD, UK}
\affiliation{Shanghai Key Laboratory of Special Artificial Microstructure and Pohl Institute of Solid State Physics and School of Physics Science and Engineering, Tongji University, Shanghai 200092, China}

\begin{abstract}
Thermally assisted magnetic writing is an important technology utilizing temperature dependent magnetic properties to enable orientation of a magnetic data storage medium. Using an atomistic spin model we study non-equilibrium field cooled magnetization processes on sub-nanosecond timescales required for device applications. We encapsulate the essential physics of the process in a TRM-T curve and show that for fast timescales heating to the Curie temperature is necessary where the magnetic relaxation time is shortest. Furthermore we demonstrate the requirement for large magnetic fields to achieve a high thermoremanent magnetization necessary for fast recording or data rates.
\end{abstract}

\pacs{75.78.-n,75.60.Jk,85.70.-w,85.75.-d,75.10.Hk}
\maketitle

The increasing requirements for digital data storage over the past 30 years has led to an exponential growth in the storage capacity of magnetic recording media, with a typical hard drive today capable of storing around 900 Gigabits of data per square inch of recording medium. Magnetic recording has seen several significant enhancements over its lifetime, from the shift to giant magnetoresistive spin valve sensors\cite{GMR}, longitudinal to perpendicular recording\cite{perp} and more recently to composite media designs such as exchange coupled composite media.\cite{ecc} At the same time technologies such as magnetic random access memory (MRAM) promise the possibility of a universal memory\cite{Akerman}. However, current perpendicular magnetic recording technology is approaching its fundamental limit due to the magnetic recording trilemma\cite{tri}, which gives the three competing requirements for conventional magnetic recording: signal to noise ratio, thermal stability and writability. Current media designs are approaching the limits of the trilemma primarily due to the limited write field available in the inductive write head which is insufficient to reorient the high anisotropy media. One solution is a technology called Heat Assisted Magnetic Recording\cite{Rottmayer:2006ej,Kryder:2008kt} (HAMR), which utilizes the temperature dependence of the magnetic properties using laser heating to enable writing of the media, while the data is stored at room temperature for thermal stability in excess of ten years. 

Unlike conventional recording, where the applied field initiates deterministic switching of the magnetic state, thermally assisted writing requires that at the writing temperature the applied field is strong enough to overcome the \textit{thermal} writability to orient the magnetization $m$, defined by\cite{Chantrell:1985tw,EvansAPL2012}
\begin{equation}\label{eq:m_inf}
m = \tanh \left(\frac{\mu_0 \mu H}{k_B T_{\mathrm{wr}}} \right)
\end{equation}
where $\mu_0$ is the permeability of free space, $\mu$ is the magnetic moment, $k_B$ is the Boltzmann constant, $H$ is the applied field and $T_{\mathrm{wr}}$ is the writing temperature. In combination with the usual requirements for the trilemma, this leads to the magnetic recording quadrilemma,\cite{EvansAPL2012} which also places a fundamental limit on the ultimate achievable recording density\cite{Richter}. For reasonably sized grains (with large $\mu$), the thermal writability is close to one, meaning that given an infinite amount of time the writing process will achieve the thermal equilibrium value of the magnetization defined in Eq.~\ref{eq:m_inf}. However, increasing data densities require that the HAMR switching process is completed on the sub-nanosecond timescale. This essentially defines HAMR as a \textit{non-equilibrium} field cooling (NEFC) process, where the thermal writability is limited by the ability to achieve thermal equilibrium during the writing process. This same physical idea is utilized in next-generation thermally assisted MRAM where the magnetization is reduced by Joule heating to enable writing of the cell on the nanosecond timescale\cite{TAMRAM}. In both cases the dynamics of the thermally assisted writing process are crucial to achieving reliable switching suitable for device applications. 

In this letter we investigate the importance of sub-nanosecond magnetization dynamics on non-equilibrium field cooling processes using atomistic simulations. We determine the effect of the cooling rate and applied field strength on the achievable thermoremanent magnetization necessary for technological applications. We find that for fast cooling rates it is necessary to heat the magnetic material close to the Curie temperature and apply large magnetic fields in order to successfully orient the magnetization.

To simulate the dynamic properties of the recording medium during the non-equilibrium field-cooling process up to and beyond the Curie temperature we utilize an atomistic spin model\cite{EvansJPCM2014} using the \textsc{vampire} software package\cite{vampire}. In this work we model material parameters for $L1_0$ FePt due to its large magnetocrystalline anisotropy and low Curie temperature of around 700 K\cite{Okamoto:2002tw}. The energetics of the system are given by the Heisenberg spin Hamiltonian 
\begin{equation}\label{C2SpinHamiltonian}
\mathcal{H} = -\sum_{i\ne j} J_{ij} \mathbf{S}_i \cdot \mathbf{S}_j -k_u \sum_i \left( S_i^z \right)^2 - \sum_i \mu_{\mathrm{s}} S_i^z H_{\mathrm{app}}^z
\end{equation}
where $J_{ij} = 3.0 \times 10^{-21}$ J/link is the nearest neighbor exchange interaction between the sites $i$ and $j$, $\mathbf{S}_i$ is the local spin moment, $\mathbf{S}_j$ is the spin moment of a neighboring atom $j$, $k_u$ is the anisotropy energy per atom, $\mu_{\mathrm{s}} = 1.5 \mu_{\mathrm{B}}$ is the effective atomic spin moment and $H_{\mathrm{app}}^z$ is the externally applied magnetic field. Both the easy axis and externally applied field are both orientated along the $z$-axis. The anisotropy has been set so that the energy barrier $K V /k_{\mathrm{B}} T = 60$ leads to a characteristic relaxation time at room temperature in excess of ten years, a typical design requirement for magnetic recording media or MRAM. The high anisotropy leads to a room temperature coercive field of around 2.2 Tesla, significantly higher than the fields available to conventional inductive write heads and thus requiring a thermal assist to orient the magnetization. The minimum write temperature is determined by the temperature dependent coercivity and for an external applied field of 0.8 Tesla is $T_{wr} \sim 600$ K, which is significantly below the Curie temperature of 690K. Consequently one would generally expect deterministic switching of the magnetization at this temperature. The system is integrated using the stochastic Landau-Lifshitz-Gilbert equation using the Heun numerical scheme\cite{EvansJPCM2014} using a time step of 0.1 fs to ensure correct simulation of the dynamics at elevated temperatures. The Gilbert damping constant is set at $\lambda=0.1$ typical of FePt\cite{FePtlambda}. The simulated system has a face-centred-cubic crystal structure and each grain is in the form of a cylinder with height and diameter of 5 nm.

\begin{figure}[!lt]
\center
\includegraphics[width=8cm, trim=15 30 15 0]{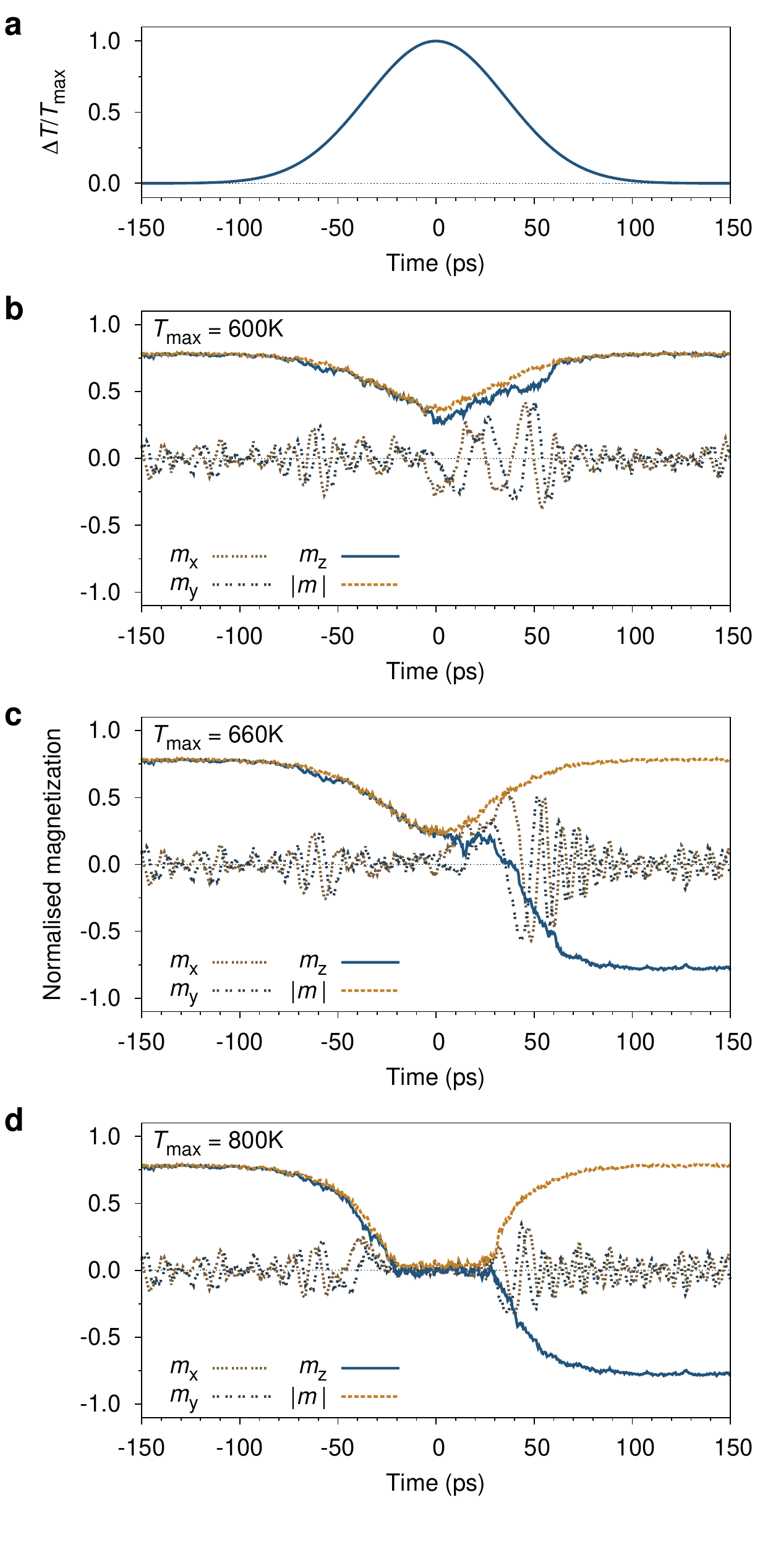}
\caption{(\textbf{a}) Normalized temporal temperature profile and (\textbf{b-d}) comparative magnetization dynamics for a single 5 nm diameter grain under the action of an applied field for different maximum temperatures, $T_{\mathrm{max}}$. (\textbf{b}) For $T_{\mathrm{max}} = 600$ K ($ \sim 0.87 T_c$), the grain demagnetizes, but remains orientated along the original direction after cooling. (\textbf{c}) For $T_{\mathrm{max}} = 660$ K ($ \sim 0.96 T_c$), the magnetization now shows thermal activation over the energy barrier via a precessional reversal mode. (\textbf{d}) For $T_{\mathrm{max}} = 800$ K (above $T_c$), the system is completely demagnetized and switching occurs via a linear reversal mode. (Color online.)}
\label{fig:Mvst}
\end{figure}

The NEFC process is simulated by applying a time-dependent temperature pulse representing heating and cooling given by 
\begin{equation}
T(t) = T_{\mathrm{rt}} + (T_{\mathrm{max}}-T_{\mathrm{rt}}) \exp\left(-\frac{t^2}{\tau_{\mathrm{p}}^2}\right)
\label{eq:pulse}
\end{equation}
where $\tau_{\mathrm{p}}$ is the pulse width, $T_{\mathrm{max}}$ is the peak temperature achieved during the pulse, and $T_{\mathrm{rt}}$ is room temperature, set at 300 K. A typical temperature-time profile is plotted in Fig.~\ref{fig:Mvst}(a) for a pulse width of $\tau_{\mathrm{p}} = 50$ ps. A constant reversing field is applied throughout the simulation along the $-z$ direction in order to attempt to orient the magnetization.

During the heating and cooling process the characteristic magnetization dynamics are strongly dependent on the peak temperature achieved during the simulation, which has a direct impact on the reliability of the NEFC process. The calculated magnetization dynamics for a single grain for different maximum temperatures and a characteristic pulse width of $\tau_{\mathrm{p}} = 50$ ps are shown in Fig.~\ref{fig:Mvst}(b-d). The dynamic simulations show that the reversal mechanism and associated timescale of the switching is critically dependent on the temperature the system reaches during the field cooling process. Elevated temperatures generally lead to faster reversal due to greater thermal fluctuations, but for precessional switching the timescale is still of the order of 100 ps. Heating the system to temperatures in the vicinity of the Curie point allows for magnetic switching via a linear reversal mode\cite{Kazantseva,Barker:2010hfa}, where longitudinal fluctuations of the magnetization are energetically favored over precession with a characteristic switching time of a picosecond.

\begin{figure}[!rt]
\center
\includegraphics[width=8cm, trim=15 10 15 0]{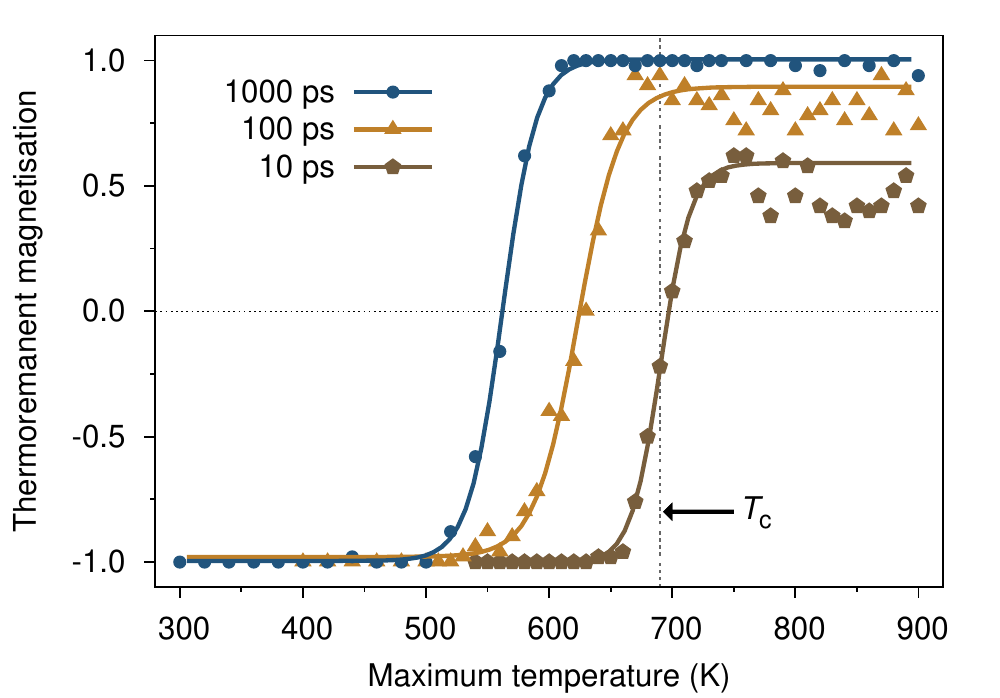}
\caption{Simulated normalized thermoremanent magnetization of a collection of 100 grains as a function of the maximum temperature reached for different characteristic heating and cooling times $\tau_{\mathrm{p}}$ for a gaussian heating/cooling profile. A magnetic field of 0.8T is applied along the -$z$ direction, opposed to the initial magnetization of the grain. Lines provide guides to the eye. The vertical dashed line indicates the Curie point. (Color online.)}
\label{fig:RPvsT}
\end{figure}

Although the single grain magnetization dynamics reveal changes in the physical reversal behavior, they provide no information about the statistical reliability of the reversal process and the achievable thermoremanent magnetization, that is the net magnetization after cooling to ambient temperature. A high thermoremanent magnetization is characteristic of a reliable switching process and essential for a usable data storage technology. To calculate the thermoremanent magnetization it is necessary to simulate an ensemble of grains which we do by simulating 100 non-interacting grains and determining the number of grains switched during the NEFC process. Calculated thermoremanent magnetization curves as a function of the maximum temperature reached during the pulse (TRM-T plot) for characteristic cooling times $\tau_{\mathrm{p}}$ of 10 ps, 100 ps, and 1000 ps are plotted in Fig.~\ref{fig:RPvsT}.

For the longest cooling time $\tau_{\mathrm{p}} = 1000$ ps, the system has sufficient time to reach thermal equilibrium during the cooling process and therefore as long as the temperature is greater than the write temperature of $T_{\mathrm{wr}} \sim 600$K almost all the grains are aligned with the field after cooling to $T_{\mathrm{rt}}$. For the intermediate cooling time of $\tau_{\mathrm{p}} = 100$ ps, the thermoremanent magnetization only approaches practical levels ($>90$\%) in the vicinity of the Curie temperature due to the presence of the linear reversal mode. At lower maximum temperatures the timescale for thermally activated reversal is too long to allow switching, even above the write temperature where the coercivity is less than the applied field. For $T_{\mathrm{max}}$ above $T_{\mathrm{c}}$ the thermoremanent magnetization reduces slightly due to a dynamic effect during the cooling process. Below a critical blocking temperature\cite{Chantrell:1985tw} $T_{\mathrm{B}}$ no reversal is permitted due to a high energy barrier. Therefore, the temporal reversal window between $T_{\mathrm{B}} < T < T_{\mathrm{c}}$ is critical to the overall thermoremanent magnetization. For the $\tau_{\mathrm{p}} = 100$ ps cooling time this essentially means that heating to temperatures above $T_{\mathrm{c}}$ is disadvantageous to maximizing the thermoremanent magnetization. More worrying is the inability of the thermoremanent magnetization to saturate close to 1, even at $T_{\mathrm{c}}$. This dynamic effect arises not due to the quadrilemma\cite{EvansAPL2012} (since for 1 ns cooling this is readily achieved), but due to freezing of the magnetization in the wrong direction when the system reaches the blocking temperature. This illustrates the importance of cooling as slowly as possible through the Curie temperature. 

For the shortest cooling time of $\tau_{\mathrm{p}} = 10$ ps no switching is seen until the peak temperature reaches the Curie temperature, illustrating the disparity in the characteristic relaxation times between the linear reversal and thermally activated precessional switching. As with the 100 ps case, the achieved thermoremanent magnetization is well below that for the 1000 ps case, owing to the freezing of the magnetization at the blocking temperature.

From an engineering perspective the TRM-T plot presented in Fig.~\ref{fig:RPvsT} is important as it encapsulates the essence of the NEFC process. It determines, for a given cooling rate, the temperature required for reliable switching of the grains. For HAMR, if the thermal gradient in the media caused by the laser is known, then the TRM-T plot also gives the allowable temperature gradient to avoid writing adjacent tracks of data, since if the temperature at the adjacent track is less than the maximum $T_{\mathrm{max}}$ for a TRM of -1, no grains will switch. In addition the temperature width of the TRM-T curve gives an estimate of the recording jitter, as the down-track thermal gradient leads to a variation of the switching probability, contributing to an effective switching-field distribution. In this sense, the ideal TRM-T curve is a $\delta$-function, transitioning from TRM=-1 to TRM = 1 at a single $T_{\mathrm{max}}$.

\begin{figure}[!t]
\center
\includegraphics[width=8cm, trim=15 10 15 0]{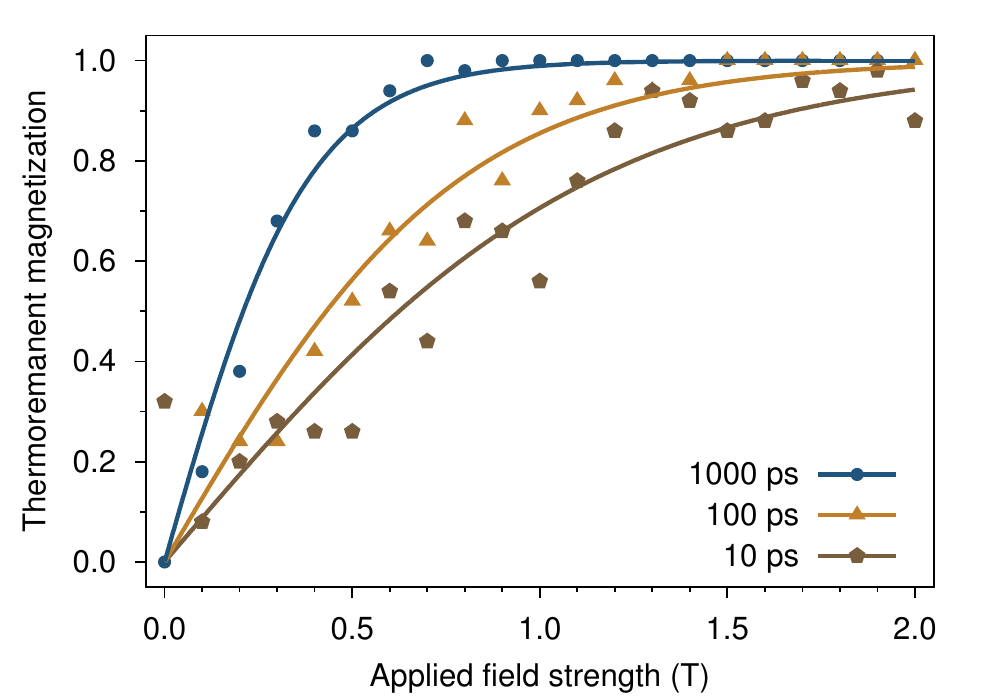}
\caption{Field dependent thermoremanent magnetization for cooling through $T_c$ at different cooling rates. Lines provide a guide to the eye. Increasing the applied field strength causes more grains to be orientated along the field direction, with slow cooling leading to progressively higher thermoremanent magnetization values. (Color online.)}
\label{fig:rp-Ht}
\end{figure}

Finally we consider the effect of the external applied field strength on the achievable dynamic thermoremanent magnetization. The same system of 100 grains is first equilibrated above $T_c$ at 800K for 10 ps (enough to reach thermal equilibrium) and then cooled exponentially with different characteristic cooling times $\tau_{\mathrm{p}}$ with a temperature-time profile given by $T(t) = T_{\mathrm{rt}} + (T_{\mathrm{max}}-T_{\mathrm{rt}}) \exp ( -t/\tau_{\mathrm{p}})$. The thermoremanent magnetization is plotted as a function of the strength of the applied field for different cooling times in Fig.~\ref{fig:rp-Ht}. 

For zero field, the thermoremanent magnetization is around 0 due to there being no applied field to break the symmetry, hence grains have an equal probability of being along or against the field axis. For the longest cooling time of 1000 ps large fields approaching 1T are required to achieve a high thermoremanent magnetization, arising from the thermal writability encapsulated by Eq.~\ref{eq:m_inf}. At faster cooling rates the reduced time to achieve thermal equilibrium in the field above the blocking temperature significantly reduces the achievable thermoremanent magnetization and much larger applied field strengths are required to achieve reliable writing of the grains. The strong field dependence of the thermoremanent magnetization causes a potential problem for ultrahigh densities or for HAMR variants such as pulsed HAMR where rapid cooling is required and must therefore be accompanied by strong head fields.

In conclusion, we have presented atomistic simulations of the NEFC process and encapsulated the core physics as a TRM-T curve. We have demonstrated the importance of slow cooling to achieve a high thermoremanent magnetization and the requirement for strong applied fields and heating to the vicinity of the Curie temperature for fast cooling rates necessary for pulsed HAMR or ultrahigh recording densities. For thermally assisted MRAM the equivalent requirement for fast switching is heating to the N\'eel temperature of the antiferromagnet\cite{TAMRAM,EBHAMR}. While strong heating is relatively straightforward to achieve, strong applied fields are much more challenging. However alternative technologies such as ultrafast optomagnetic recording\cite{lambert2014,EvansSFiM2014} or composite media designs\cite{Suess} utilizing the exchange field to assist the writing process hold significant promise to enable order-of-magnitude improvements in data storage capacity at high data rates.

This work was supported by the European Community's Seventh Framework Programme (FP7/2007-2013) under grant agreement 281043 (\textsc{femtospin}). The authors would like to thank Pierre Asselin, Ondrej Hovorka, Roy Chantrell and Pin-Wei Huang for helpful discussions.

\end{document}